 \documentclass[sigconf]{acmart}
 \usepackage[most]{tcolorbox}
\AtBeginDocument{%
  \providecommand\BibTeX{{%
    \normalfont B\kern-0.5em{\scshape i\kern-0.25em b}\kern-0.8em\TeX}}}

\setcopyright{none}

\begin{document}
\setcopyright{none}
\settopmatter{printacmref=false} 
\renewcommand\footnotetextcopyrightpermission[1]{} 
\pagestyle{plain} 

\title{On the Energy  Consumption of Different Dataframe Processing Libraries - An Exploratory Study}

\author{Shriram Shanbhag, Sridhar Chimalakonda}
\affiliation{\textit{Research in Intelligent Software \& Human Analytics (RISHA) Lab}\\
Department of Computer Science \& Engineering \\
Indian Institute of Technology Tirupati
\country{India}\\}
\email{{cs20s503, ch}@iittp.ac.in}
\renewcommand{\shortauthors}{Shanbhag et al.}

\begin{abstract}
\textbf{\textit{Background}}: The energy consumption of machine learning and its impact on the environment has made energy efficient ML an emerging area of research. However, most of the attention stays focused on the model creation and the training and inferencing phase. Data oriented stages like preprocessing, cleaning and exploratory analysis form a critical part of the machine learning workflow. However, the energy efficiency of these stages have gained little attention from the researchers. \textbf{\textit{Aim}}: Our study aims to explore the energy consumption of different dataframe processing libraries as a first step towards studying the energy efficiency of the data oriented stages of the machine learning pipeline. \textbf{\textit{Method}}: We measure the energy consumption of 3 popular libraries used to work with dataframes, namely \textit{Pandas}, \textit{Vaex} and \textit{Dask} for 21 different operations grouped under 4 categories on 2 datasets. \textbf{\textit{Results}}: The results of our analysis show that for a given dataframe processing operation, the choice of library can indeed influence the energy consumption with some libraries consuming 202 times lesser energy over others.  \textbf{\textit{Conclusion}}: The results of our study indicates that there is a potential for optimizing the energy consumption of the data oriented stages of the machine learning pipeline and further research is needed in the direction. 
\end{abstract}





\maketitle

\section{Introduction}
Machine learning is a branch of artificial intelligence that focuses on using data and algorithms to make the systems learn. The problems addressed by machine learning applications include computer vision \cite{krizhevsky2012imagenet}, speech recognition \cite{amodei2016deep}, natural language processing \cite{cai2017encoder}, machine translation \cite{wu2016google} and software engineering \cite{lee2017applying} among others. Data plays a critical role in machine learning as it is required to train the models to make predictions. In recent years, there is a drastic improvement in performance of machine learning models driven by new architectures like neural networks, availability of GPU based powerful hardware infrastructure and huge amounts of data to train the machine learning models.However, there is also a critical and growing concern of energy consumption of these models. A study from 2019 by Strubell et al.  \cite{strubell2019energy} revealed that a single neural network model with 626,155 parameters has a carbon footprint equivalent to lifetime carbon footprint of 5 cars. 


The machine learning workflow \cite{amershi2019software} typically contains “data oriented” stages like data preprocessing, cleaning and exploratory data analysis. The use of relational databases for these stages come with a lot of limitations. Oftentimes, the data collected would not be well structured making it difficult to work with database queries \cite{wu2020dataframe}. Writing complex queries would require a strong familiarity with the schema which is not easy in case of wide tables with many columns \cite{gathani2020debugging}. Dataframes have been introduced as an alternative to overcome some of these limitations. Dataframe is a data structure that organizes data into rows and columns like a spreadsheet. Over the years, the use of dataframes has become widely popular as they provide an intuitive way of storing and working with data \cite{wu2020dataframe, petersohn2021dataframe}. They are convenient for use in REPL style imperative interfaces and data science notebooks \cite{perez2015project}.  Several libraries \cite{mckinney2011pandas, rocklin2015dask, breddels2018vaex} have been developed to enable the handling and manipulation of dataframes. These libraries provide a convenient interface to work with dataframes and are widely used for exploratory analysis, data ingest, data preparation and feature engineering. Dataframes offer several features including implicit ordering on both rows and columns by treating them symmetrically. The dataframe manipulation libraries offer data analysis modalities that include relational operators like filter, join, transpose and spreadsheet-like operators such as pivot.

The dataframe manipulation libraries are used in the data oriented stages of the machine learning workflow. \textit{Pandas}, the most popular dataframe processing library for instance, has been downloaded over 300 million times\footnote{\url{https://pypistats.org/packages/pandas}} with its Github repository starred over 33k times as of April 2022\footnote{\url{https://github.com/pandas-dev/pandas}}. Machine learning however, has become very data intensive in recent years. As the  size of the data gets bigger, it cannot fit in the system RAM leading to the need for more efficient, out-of-core dataframe processing methods.This has prompted the need for development of faster and scalable dataframe processing methods. This has led to the development of  scalable ways of dataframe processing using the libraries such as \textit{Dask}, \textit{Vaex}, RAPIDS etc. These libraries enable scaling either through parallelization or through the use of efficient computation techniques and memory usage. While the emphasis in these libraries remains focused on achieving faster processing and scalability, energy consumption aspect of these libraries is typically ignored. 

In the recent years, energy consumption of the software systems have gained a major attention from the researchers. Several works have observed a link between design of a software and its energy consumption \cite{hindle2014greenminer, cruz2019catalog}. The lack of knowledge about the energy implications of poor design choices have also been observed in a study \cite{pinto2017energy}. In the domain of machine learning, the computation and energy demands have been increasing exponentially which would require a larger energy production. This has the potential to cause a significant portion of the carbon emissions \cite{edenhofer2015climate}. Owing to this many researchers have focused on energy efficiency of machine learning. There has been a focus on measuring the energy consumption of machine learning models \cite{lee2017lognet, sanh2019distilbert}, reducing the energy requirements of training \cite{jiao2018energy, sarwar2018energy} and inferencing \cite{sanh2019distilbert}. However, the work on energy efficient machine learning have largely ignored the data oriented stages of the machine learning pipeline. To fill this gap, as a first step, we perform an exploratory study on energy consumption of dataframe libraries. 

In this paper, we perform an empirical evaluation of the energy consumption of three popular libraries namely \textit{Pandas}\footnote{https://pandas.pydata.org/}, \textit{Vaex}\footnote{https://vaex.io} and \textit{Dask}\footnote{https://docs.dask.org/en/stable/dataframe.html} for four categories of tasks. Although \textit{Dask} is not a library specifically mean for working with dataframes, it does contain a dataframe module that can be perform operations on dataframes efficiently using computation graphs \cite{rocklin2015dask}. The four types of tasks include input-output operations, handling missing data, row/column operations and statistical aggregation. 

The rest of the paper is organized as follows. We provide a brief description about the libraries used in our study in Section \ref{sec:libraries}. The tasks addressed in the study are discussed in Section \ref{sec:tasks}. The datasets used in the study are presented in Section \ref{sec:datasets} The experimental settings and the procedure followed are described in Section \ref{sec:procedure}. The results of the experiments are presented in Section \ref{sec:results}. Section \ref{sec:threats} discusses the potential threats that may impact the validity of our study. The discussion, related work and conclusion are presented in Section \ref{sec:discussion}, Section \ref{sec:related_work} and Section \ref{sec:conclusion_future} respectively. 
\begin{figure}
    \centering
    \includegraphics[scale=0.5]{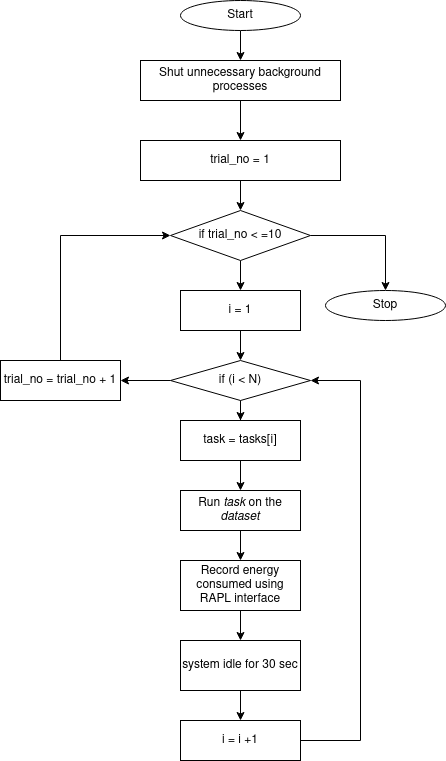}
    \caption{Flowchart describing the experimental procedure followed in the study based on \citet{georgiou2022green}}
    \label{fig:flowchart}
\end{figure}
\section{Dataframe Processing Libraries}
\label{sec:libraries}
The data used in machine learning is normally structured in tabular form called dataframes. Dataframe is a 2 dimensional labeled data structure with columns of potentially different types. Dataframes provide a common structure to work with data from domains such as finance, statistics, social sciences among many others. Libraries like \textit{Pandas} and \textit{Vaex} provide a rich set of functionalities to manipulate the dataframes that enable easier data cleaning and preparation.

\textit{Pandas} \cite{mckinney2011pandas} is an open source Python library for data manipulation developed in the year 2008. It provides data structures and operations for manipulation of tabular data. It was originally developed by \textit{Wes McKinney} for working with financial data. In addition to data manipulation, it can also be used for data analysis and visualization. \textit{Vaex} \cite{breddels2018vaex} on the other hand is a Python library for working with lazy out of core dataframes on huge amounts of tabular data. The project has over 7K stars and over 500 forks on GitHub as of March 2022\footnote{\url{https://github.com/vaexio/vaex}}. \textit{Vaex} claims to be an alternative to \textit{Pandas} that has a more efficient and faster implementation of data manipulation functions using memory mapping, a zero memory copy policy, and lazy computations. 

\textit{Dask} is an opens source library for parallel computing in Python. \textit{Dask} provides support for dataframe processing through a module that has an interface similar to \textit{Pandas}. The module implements a large dataframe out of many \textit{Pandas} dataframes \cite{rocklin2015dask}. It uses a threaded scheduler that enables efficient computation on partitioned datasets \cite{rocklin2015dask}. \textit{Dask} minimizes the amount of data held in memory while parallelizing the tasks using computation graphs \cite{rocklin2015dask}. 

\section{Addressed Tasks}
\label{sec:tasks}
In order to compare the energy consumption of these libraries while working with the dataframes, we used different libraries to perform the same set of operations on three different datasets. We picked some of the commonly performed operations on dataframes as  tasks for our evaluation. The operations included loading of the csv file, dropping columns, conatenatination of dataframes, duplicate removal, merging dataframes, subset, sampling, and aggregation operations such as sum, median, min, max, unique. A brief description of each of these tasks is given in Table \ref{tab:tasks}.

Dataframe processing libraries can be used to perform several operations on dataframes including read/write operations, visualization, relational and statistical operations. In order to compare the energy consumption of these libraries while working with the dataframes, we used different libraries to perform the same set of operations on three different datasets. we address a few frequently performed tasks input/output operations, relational operation and statistical operations. 

\subsection{Input/Output Operations}
Dataframes can be read and stored in several formats including CSV, HDF5, Parquet, JSON etc. The choice of the format is typically decided based on the type of data being stored and the type of operations being performed on the data. As an example, storing dataframes in a format like Parquet can be benifitial when used in case of big data systems like Hadoop or Spark. However, unlike CSV, the storage format is also not human readable. Since our work focuses on energy, we compare the energy consumption of input and output operations on different file formats used to represent the dataframes. Specifically we compare the energy consumption values of reading and writing the dataframes in three popular formats namely CSV, JSON and HDF5 using \textit{Pandas}, \textit{Vaex} and \textit{Dask} libraries. 

CSV stands for \textit{comma-separated values}. CSVs store data as plain text where each row is a line of columns separated by ",". It is very widely used due to the fact that its human readable and can be read from and written to by most data software. JSON stands for \textit{JavaScript Object Notation} is another human readable format that uses text to store data using attribute-value pairs and arrays. Like CSV,  JSON can also be readily displayed and edited in simple editors. It also has an advantage of being a preferred format for transmitting data in web applications. HDF5 stands for \textit{Hierarchical Data Format 5} suited for storing and organizing large amounts of heterogeneous data. The data is stored as an internal file-like structure which provides the ability to randomly access different parts of the data. HDF5 offers efficient storage and faster access when dealing with huge data. 

\subsection{Handling missing data}
Missing data is a very common problem in real-world datasets. They are generally referred to as NA values in dataframes. Missing data is generally caused by lack of collection of existing data or because of the non-existence of the data. Missing data in the dataframes is undesirable as they generally interfere with the operations and calculations being performed on rows/columns. This makes handling of missing data a very frequently performed task in data preprocessing. The missing data handling functions compared in our study are found in Table \ref{tab:tasks}

\subsection{Row/Column operations}
Dataframes represent the data in the form of a table consisting of rows and columns. Dataframes support relational operations like join, filter, groupby etc. They also support reshape and transpose operations on the tables. This makes it easier to merge data obtained from multiple sources in multiple dataframes and order them in a desired way to enable  a better exploratory analysis. They also enable the users to apply functions on rows/columns that allow modification and filtering of the data. The missing data handling functions compared in our study are found in Table \ref{tab:tasks}

\subsection{Statistical Aggregation Operations}
Statistical aggregation functions enable the users to perform statistical analysis of their data. They can be used for operations like aggregating multiple values under a row or column using a single summary value, computing the correlation and covariance between pairs of rows/columns, finding the unique occurrences, count of rows/columns etc. These functions are frequently used in exploratory data analysis to get the summary or the description of the dataset. The statistical aggregation functions compared in our study are found in Table \ref{tab:tasks}

\begin{table*}[]
\begin{tabular}{|l|l|l|l|}
\hline
\textbf{Task ID} & \textbf{Task Name}             & \textbf{Description}                                                   & \textbf{Category}                \\ \hline
csvin   & Read CSV              & Read a comma-separated values (csv) file into dataframe       & I/O operation           \\ \hline
jsonin  & Read JSON             & Read JSON string into a Dataframe                             & I/O operation           \\ \hline
hdfin   & Read HDF5             & Read an heirarchical data format 5 (hdf5) file into dataframe & I/O operation           \\ \hline
csvout  & Write to CSV          & Write dataframe into a comma-separated values (csv) file      & I/O operation           \\ \hline
jsonout & Write to JSON         & Write dataframe into a JSON file                              & I/O operation           \\ \hline
hdfout  & Write to HDF5         & Write dataframe into a heirarchical data format 5 (hdf5) file & Handling missing data   \\ \hline
isna    & Detect missing values & Detect missing values in a given series object                & Handling missing data   \\ \hline
dropna  & Drop missing values   & Drop rows/columns containing missing values                   & Handling missing data   \\ \hline
fillna  & Fill missing values   & Fill missing values with a default                            & Handling missing data   \\ \hline
replace & Replace values        & Replace a value with a specified value                        & Handling missing data   \\ \hline
drop    & Drop rows/columns     & Drop specified labels from rows or columns                    & Row/Column operations        \\ \hline
groupby & Groupby Operation     & Group rows that have the same values together                 & Row/Column operations        \\ \hline
concat  & Concatenation         & Concatenate dataframes along a particular axis                & Row/Column operations        \\ \hline
sort    & Sort values           & Sort the values along an axis                                 & Row/Column operations        \\ \hline
merge   & Merge dataframes      & Merge dataframes with a database-join style                   & Row/Column operations        \\ \hline
count   & Count cells           & Count the number of non-NA cells along an axis                & Statistical Aggregation \\ \hline
sum     & Sum values            & Sum values along an axis                                      & Statistical Aggregation \\ \hline
mean    & Average values        & Average values along an axis                                  & Statistical Aggregation \\ \hline
min     & Minimum value         & Find the minimum of the values along an axis                  & Statistical Aggregation \\ \hline
max     & Maximum value         & Find the maximum of the values along an axis                  & Statistical Aggregation \\ \hline
unique  & Unique values         & Find the set of unique values along the axis                  & Statistical Aggregation \\ \hline
\end{tabular}
\caption{Tasks addressed in the study}
\label{tab:tasks}
\end{table*}

\section{Datasets Used}
\label{sec:datasets}
In order to compare the energy consumption of the dataframe processing libraries, we have to perform the same set of operations on the same datasets. Towards this end, we relied on well known standard datasets from UCI Machine learning repository\footnote{\url{https://archive.ics.uci.edu/ml/index.php}}. The energy consumption of the libraries may also vary based on the size of the dataframe they process on. A library that may be more energy efficient on a smaller dataframe may become less energy efficient as the dataframes get bigger. In order to account for these variations, we performed our experiments on 2 different datasets of varying sizes from the UCI repository shown in Table \ref{tab:datasets}. The datasets used are briefly described below. 
\subsection{Adult Dataset} 
The dataset was extracted based on the 1994 census database. The task associated with the dataset is the classification task to determine whether a person makes over 50K a year or not. The dataset contains 48842 data points with 14 attributes. We would refer to this dataset as \textit{D1} in the rest of the paper. 

\subsection{Drug Review Dataset}
This dataset contains patient reviews on specific drugs along with related health conditions. It also contains the 10 star patient rating of the drug. The dataset has 215063 data points with 6 attributes. We would refer to this dataset as \textit{D2} in the rest of the paper. 

\begin{table*}[]
\begin{tabular}{|l|l|l|l|l|}
\hline
\textbf{Dataset Name} & \textbf{Characteristics}                                      & \textbf{Associated Tasks}                                                        & \textbf{No. of Data Points} & \textbf{No. of Attributes} \\ \hline
Adult                 & Multivariate                                                  & Classification                                                                   & 48842                       & 14                         \\ \hline
Durg Review           & \begin{tabular}[c]{@{}l@{}}Multivariate, \\ Text\end{tabular} & \begin{tabular}[c]{@{}l@{}}Classification,\\ Regression, Clustering\end{tabular} & 215063                      & 6                          \\ \hline
\end{tabular}
\caption{Datasets used for the study}
\label{tab:datasets}
\end{table*}

\section{Experimental Procedure}
\label{sec:procedure}
All the experiments regarding energy consumption are run on a system with an intel i5 4200 four core processor equipped with a frequency of 1.6 GHz. The system was equipped with a RAM of 8 GB. We used Python 3.8 as the programming language for all our experiments. The system had Ubuntu 20.04 installed as the operating system. We used \textit{Pandas} version 1.2.4, \textit{Vaex} version 4.8.0 and \textit{Dask} version 2022.2.1 for our study. 

To measure the energy consumption values, we used Intel's Running Average Power Limit (RAPL) interface. Using the RAPL tool is one of the most accurate ways to measure the global energy consumption of the processor \cite{david2010rapl}. Several studies related to measuring the energy consumption of the software systems have relied on the RAPL interface \cite{pereira2017energy, ournani2021comparing, kumar2019energy}. The tool provides energy consumption values of various power domains such as package, core, uncore, power plane, DRAM etc. The energy consumption values of the following domains were noted down during the experiments

\begin{itemize}
    \item Package (PKG): This domain provides the energy consumption values of all the cores, integrated graphics and uncore components such as last level caches and memory controllers.
    \item Core: The energy consumption values of all the CPU cores are provided by this domain. 
    \item Uncore: This domain provides the energy consumption values of all the caches, integrated graphics and the memory controllers.
    \item DRAM: The energy consumption values corresponding to the random access memory attached to the memory controller is provided by this domin. 
\end{itemize}

Since the PKG values are inclusive of core and non-core values, we only report the PKG values in the results in Section \ref{sec:results}. 

The experimental procedure using a single library is represented as a flowchart in Figure \ref{fig:flowchart}. At the start of the experiment, we shut down all the processes that were not required to run the OS. This was done to ensure minimal variation in the readings due to interference from the processes running in the background. 

Power consumption measurement in the system can be influenced by noise. So one of the methods used to minimize the effect of the noise on the measurements is to run the same experiment multiple times and take the mean value of the measurements for all the experiments as followed in a study by Georgiou et al. \cite{georgiou2022green}. Based on this, we ran the experiment for each operation 10 times and recorded the mean value of the energy measurements as shown in Figure \ref{fig:flowchart}. After running each task on a dataset, we let the system remain idle for 30 seconds using the command “sleep”. This was done in order to avoid the power tail states \cite{bornholt2012model} influencing the reading and allow the system to reach a stable state again before the start of another task. At the end of each task, the RAPL readings were stored in an excel sheet for further analysis. The information included the time required to run each task along with the energy consumption values of the components. 

The experiment was performed using all the three libraries considered for the study. The mean values of all the trials for each task was computed and noted down using automated python script. The scripts used for the study  along with the detailed energy consumption values of the tasks for each trial is available here\footnote{\url{https://anonymous.4open.science/r/esem2022-5D62}}.

\section{Results}
\label{sec:results}
This section discusses the results of the experiments measuring energy consumption of 3 different libraries on various dataframe operations. The energy consumption measurements for input/output operations are shown in Table \ref{tab:input} and Table \ref{tab:output}. The energy consumption of operations on handling missing data is shown in Table \ref{tab:missing_data}. The energy consumption values of table and statistical operations are shown in Table \ref{tab:table_operations} and Table \ref{tab:statistical_operations} respectively.   We use the abbreviation \textit{PKG} for the energy consumption values of all core and non-core components of the processor. The RAPL interface also provides the energy consumption of the core and non-core components separately. Since those are covered under \textit{PKG}, we do not report them explicitly. We use \textit{RAM} for energy consumption of the main memory. The RAPL interface provides the energy consumption values in micro-joules (1 micro-joule = $10^{-6}$ Joules). However, we present them in the results by converting them into milli-joules (1 mini-joule = $10^{-3}$ Joules) and rounding them off to one decimal. This is done to ensure better readability. The energy consumption values in all the tables are mean values from 10 trials of each operation performed as described in Section \ref{sec:procedure}. 

\subsection{Input/Output Operations}
\subsubsection{Input Operations}
The energy consumption of the dataframe input operation in CSV, JSON and HDF5 formats for the datasets \textit{D1} and \textit{D2} are shown in Table \ref{tab:input}. For the dataset \textit{D1}, \textit{Vaex} consumes the highest amount of energy for reading the dataframes in all three file formats. \textit{Dask} consumes the least amount of energy for inputs in CSV and HDF5 formats while \textit{Pandas} consumed the least energy for reading the dataframe in JSON format. For the dataset \textit{D2}, we observe that  \textit{Vaex} consumes the most energy for inputs in CSV and JSON formats while \textit{Pandas} consumes the most energy for input in HDF5. However, unlike \textit{D1}, \textit{Dask} consumes lesser energy than \textit{Pandas} for input operations in JSON format as well along with CSV and HDF5.However, we could not get values for HDF5 for input and output operations for \textit{Dask} library in the Drugs Review Dataset due to technical glitches because of the size of the data. 

Across all libraries and datasets, we can also observe that the reading the dataframe stored as JSON is most expensive in terms of energy. For \textit{Pandas}, while reading \textit{D1}, the energy consumption remains comparable for CSV and HDF5 formats. However, as the size of the dataset increases, HDF5 becomes less expensive. This can be observed from the values for D2 with \textit{Pandas} library in Table \ref{tab:input}. We can also observe that the energy consumption of CSV inputs for \textit{Dask} is not affected much by increase in the dataset size. The reason could be because of the lazy execution \cite{rocklin2015dask} using computation graphs where it only loads information about the header and the datatypes \cite{rocklin2015dask}.

\subsubsection{Output Operations}
The energy consumption of the dataframe output operation in CSV, JSON and HDF5 formats for the datasets \textit{D1} and \textit{D2} are shown in Table \ref{tab:output}. For dataset \textit{D1}, \textit{Dask} consumes the most energy while \textit{Pandas} consumes the least energy for outputs in all three formats. A similar observation can be made for the dataset \textit{D2}. The output of dataframe into a CSV file consumes the most energy across \textit{Pandas} and \textit{Vaex}. However, for output using \textit{Dask} library, HDF5 is the most energy-expensive file format.  

The following are some of the key highlights from the observation of the results of input/output operations.   
\begin{tcolorbox}[colback=red!5!white,colframe=red!75!black]
  \begin{enumerate}
      \item Overall, \textit{Vaex} is the most energy-expensive library for input operations. 
      \item JSON is the most expensive of the three file formats for dataframe input. 
      \item The energy consumption of a CSV input remains unaffected by the size of the dataset for \textit{Dask}. 
      \item \textit{Dask} consumes the most energy for output operations while \textit{Pandas} consumes the least. 
      \item CSV is the most energy-expensive file format for outputting a dataframe in \textit{Pandas} and \textit{Vaex}. 
      \item HDF5 is the most energy-expensive file format for output using \textit{Dask} library. 
    \end{enumerate}
\end{tcolorbox}
\begin{table*}[]
\begin{tabular}{|lllllll|llllll|}
\hline
\multicolumn{7}{|c|}{Adult Dataset}                                                                                                                                                                                                      & \multicolumn{6}{c|}{Drugs Review Dataset}                                                                                                                             \\ \hline
\multicolumn{1}{|l|}{}                                                       & \multicolumn{1}{l|}{Pandas} & \multicolumn{1}{l|}{}      & \multicolumn{1}{l|}{Vaex}   & \multicolumn{1}{l|}{}      & \multicolumn{1}{l|}{Dask}   &       & \multicolumn{2}{l|}{Pandas}                                  & \multicolumn{2}{l|}{Vaex}                                    & \multicolumn{2}{l|}{Dask}               \\ \hline
\multicolumn{1}{|l|}{\begin{tabular}[c]{@{}l@{}}File \\ format\end{tabular}} & \multicolumn{1}{l|}{PKG}    & \multicolumn{1}{l|}{DRAM}  & \multicolumn{1}{l|}{PKG}    & \multicolumn{1}{l|}{DRAM}  & \multicolumn{1}{l|}{PKG}    & DRAM  & \multicolumn{1}{l|}{PKG}      & \multicolumn{1}{l|}{DRAM}    & \multicolumn{1}{l|}{PKG}      & \multicolumn{1}{l|}{DRAM}    & \multicolumn{1}{l|}{PKG}      & DRAM    \\ \hline
\multicolumn{1}{|l|}{CSV}                                                    & \multicolumn{1}{l|}{551.2}  & \multicolumn{1}{l|}{62.5}  & \multicolumn{1}{l|}{1030.4} & \multicolumn{1}{l|}{95.1}  & \multicolumn{1}{l|}{62.6}   & 6     & \multicolumn{1}{l|}{9696.6}   & \multicolumn{1}{l|}{963.7}   & \multicolumn{1}{l|}{13042.2}  & \multicolumn{1}{l|}{1327.6}  & \multicolumn{1}{l|}{65.1}     & 6.1     \\ \hline
\multicolumn{1}{|l|}{JSON}                                                   & \multicolumn{1}{l|}{3178.7} & \multicolumn{1}{l|}{432.4} & \multicolumn{1}{l|}{3697.8} & \multicolumn{1}{l|}{480.1} & \multicolumn{1}{l|}{3471.9} & 465.2 & \multicolumn{1}{l|}{156434.5} & \multicolumn{1}{l|}{11384.1} & \multicolumn{1}{l|}{160102.2} & \multicolumn{1}{l|}{11791.9} & \multicolumn{1}{l|}{154388.9} & 11382.7 \\ \hline
\multicolumn{1}{|l|}{HDF5}                                                   & \multicolumn{1}{l|}{565}    & \multicolumn{1}{l|}{55.7}  & \multicolumn{1}{l|}{3379.2} & \multicolumn{1}{l|}{283.1} & \multicolumn{1}{l|}{502.7}  & 48.8  & \multicolumn{1}{l|}{5568.2}   & \multicolumn{1}{l|}{804.7}   & \multicolumn{1}{l|}{4272.1}   & \multicolumn{1}{l|}{418.2}   & \multicolumn{1}{l|}{NA}         &      NA   \\  \hline
\end{tabular}
\caption{Mean values for energy consumption for dataframe input in different formats. }
\label{tab:input}
\end{table*}

\begin{table*}[]
\begin{tabular}{|lllllll|llllll|}
\hline
\multicolumn{7}{|c|}{Adult Dataset}                                                                                                                                                                                                      & \multicolumn{6}{c|}{Drugs Review Dataset}                                                                                                                         \\ \hline
\multicolumn{1}{|l|}{}                                                       & \multicolumn{1}{l|}{Pandas} & \multicolumn{1}{l|}{}      & \multicolumn{1}{l|}{Vaex}   & \multicolumn{1}{l|}{}      & \multicolumn{1}{l|}{Dask}   &       & \multicolumn{2}{l|}{Pandas}                                & \multicolumn{2}{l|}{Vaex}                                  & \multicolumn{2}{l|}{Dask}               \\ \hline
\multicolumn{1}{|l|}{\begin{tabular}[c]{@{}l@{}}File \\ format\end{tabular}} & \multicolumn{1}{l|}{PKG}    & \multicolumn{1}{l|}{DRAM}  & \multicolumn{1}{l|}{PKG}    & \multicolumn{1}{l|}{DRAM}  & \multicolumn{1}{l|}{PKG}    & DRAM  & \multicolumn{1}{l|}{PKG}     & \multicolumn{1}{l|}{DRAM}   & \multicolumn{1}{l|}{PKG}     & \multicolumn{1}{l|}{DRAM}   & \multicolumn{1}{l|}{PKG}      & DRAM    \\ \hline
\multicolumn{1}{|l|}{CSV}                                                    & \multicolumn{1}{l|}{1673.1} & \multicolumn{1}{l|}{113.6} & \multicolumn{1}{l|}{2261.6} & \multicolumn{1}{l|}{260}   & \multicolumn{1}{l|}{4473.8} & 460.2 & \multicolumn{1}{l|}{28941.1} & \multicolumn{1}{l|}{2229}   & \multicolumn{1}{l|}{29620.6} & \multicolumn{1}{l|}{2688.4} & \multicolumn{1}{l|}{193428.6} & 14391.7 \\ \hline
\multicolumn{1}{|l|}{JSON}                                                   & \multicolumn{1}{l|}{774.1}  & \multicolumn{1}{l|}{76.5}  & \multicolumn{1}{l|}{1515.7} & \multicolumn{1}{l|}{206.9} & \multicolumn{1}{l|}{3683.7} & 416.8 & \multicolumn{1}{l|}{9797.8}  & \multicolumn{1}{l|}{1105.3} & \multicolumn{1}{l|}{12402.3} & \multicolumn{1}{l|}{1600.6} & \multicolumn{1}{l|}{170093.8} & 13343   \\ \hline
\multicolumn{1}{|l|}{HDF5}                                                   & \multicolumn{1}{l|}{288.7}  & \multicolumn{1}{l|}{34.9}  & \multicolumn{1}{l|}{1713.6} & \multicolumn{1}{l|}{138.8} & \multicolumn{1}{l|}{5336.7} & 684.1 & \multicolumn{1}{l|}{3993.2}  & \multicolumn{1}{l|}{523.8}  & \multicolumn{1}{l|}{4732.7}  & \multicolumn{1}{l|}{611.1}  & \multicolumn{1}{l|}{NA}         &  NA       \\ \hline
\end{tabular}
\caption{Mean values for energy consumption for dataframe output in different formats. }
\label{tab:output}
\end{table*}

\subsection{Handling Missing Data}
\begin{figure}
    \centering
    \includegraphics[scale=0.4]{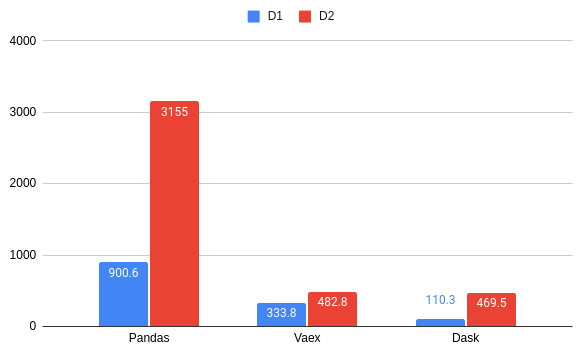}
    \caption{Cumulative energy consumption of missing data operations of the libraries for both the datasets}
    \label{fig:md}
\end{figure}
The energy consumption values of the missing data handling functions are shown in Table \ref{tab:missing_data}. For the dataset \textit{D1}, \textit{Pandas} consumes the highest energy for \textit{isna}, \textit{dropna}, and \textit{fillna} operations. While \textit{Dask} consumes the highest energy for \textit{replace} operation, it is very close the energy consumption of \textit{Pandas} with a difference of just 1.4 milli-joules. \textit{Vaex} consumes the least energy for \textit{isna} and \textit{replace} while \textit{Dask} consumes the least energy for \textit{dropna} and \textit{fillna} operations. 

For the dataset \textit{D2}, \textit{Pandas} consumes the most energy for all operations. \textit{Vaex} consumes the lowest energy for alloperations except \textit{fillna}. \textit{Dask} is the least expensive library for \textit{fillna} operation. It must be noted that for \textit{isna} and \textit{replace} functions for both datasets, the energy values of both \textit{Dask} and \textit{Vaex} are comparable with a maximum difference between them being just 5.6 milli-joues. 
The cumulative energy consumption of missing data operations of the libraries for both the datasets is shown in Figure \ref{fig:md}.

The following are some of the key highlights from the results on missing data handling operations. 
\begin{tcolorbox}[colback=red!5!white,colframe=red!75!black]
  \begin{enumerate}
     \item Overall, \textit{Pandas} is the most energy-expensive library for missing data handling operations. 
     \item \textit{Dask} is the most energy-efficient library for \textit{fillna} operation on both the datasets. 
     \item \textit{Vaex} is the most energy-efficient library for \textit{isna} and \textit{replace} operations on both datasets.
    \end{enumerate}
\end{tcolorbox}

\begin{table*}[]
\begin{tabular}{|lllllll|llllll|}
\hline
\multicolumn{7}{|c|}{Adult Dataset}                                                                                                                                                   & \multicolumn{6}{c|}{Drugs Review Dataset}                                                                                                            \\ \hline
\multicolumn{1}{|l|}{}          & \multicolumn{2}{l|}{Pandas}     & \multicolumn{2}{l|}{Vaex}   & \multicolumn{2}{l|}{Dask}      & \multicolumn{2}{l|}{Pandas}                              & \multicolumn{2}{l|}{Vaex}                             & \multicolumn{2}{l|}{Dask}         \\ \hline
\multicolumn{1}{|l|}{Operation} & \multicolumn{1}{l|}{PKG}    & \multicolumn{1}{l|}{DRAM} & \multicolumn{1}{l|}{PKG}   & \multicolumn{1}{l|}{DRAM} & \multicolumn{1}{l|}{PKG}  & DRAM & \multicolumn{1}{l|}{PKG}    & \multicolumn{1}{l|}{DRAM}  & \multicolumn{1}{l|}{PKG}  & \multicolumn{1}{l|}{DRAM} & \multicolumn{1}{l|}{PKG}   & DRAM \\ \hline
\multicolumn{1}{|l|}{isna}      & \multicolumn{1}{l|}{26.2}   & \multicolumn{1}{l|}{2.3}  & \multicolumn{1}{l|}{9.5}   & \multicolumn{1}{l|}{0.9}  & \multicolumn{1}{l|}{10.6} & 1    & \multicolumn{1}{l|}{184.6}  & \multicolumn{1}{l|}{21.6}  & \multicolumn{1}{l|}{10}   & \multicolumn{1}{l|}{0.7}  & \multicolumn{1}{l|}{10.6}  & 0.9  \\ \hline
\multicolumn{1}{|l|}{dropna}    & \multicolumn{1}{l|}{392.8}  & \multicolumn{1}{l|}{26.3} & \multicolumn{1}{l|}{137.7} & \multicolumn{1}{l|}{13.8} & \multicolumn{1}{l|}{39}   & 3    & \multicolumn{1}{l|}{1329.9} & \multicolumn{1}{l|}{111.1} & \multicolumn{1}{l|}{77}   & \multicolumn{1}{l|}{7.9}  & \multicolumn{1}{l|}{324.5} & 45.3 \\ \hline
\multicolumn{1}{|l|}{fillna}    & \multicolumn{1}{l|}{411.5}  & \multicolumn{1}{l|}{29.3} & \multicolumn{1}{l|}{155.7} & \multicolumn{1}{l|}{10.4} & \multicolumn{1}{l|}{41}   & 3    & \multicolumn{1}{l|}{1263}   & \multicolumn{1}{l|}{99.7}  & \multicolumn{1}{l|}{63.4} & \multicolumn{1}{l|}{4.4}  & \multicolumn{1}{l|}{34.6}  & 3.1  \\ \hline
\multicolumn{1}{|l|}{replace}   & \multicolumn{1}{l|}{11.2}   & \multicolumn{1}{l|}{1}    & \multicolumn{1}{l|}{5.4}   & \multicolumn{1}{l|}{0.4}  & \multicolumn{1}{l|}{11.8} & 0.9  & \multicolumn{1}{l|}{124.7}  & \multicolumn{1}{l|}{20}    & \multicolumn{1}{l|}{4.6}  & \multicolumn{1}{l|}{0.4}  & \multicolumn{1}{l|}{9.9}   & 0.7  \\ \hline
\end{tabular}
\caption{Mean values for energy consumption for missing data handling operations. }
\label{tab:missing_data}
\end{table*}
\subsection{Row/Column Operations}
The energy consumption values for the row/column operations are shown in Table \ref{tab:table_operations}. For dataset \textit{D1}, \textit{Vaex} is the least energy-expensive library on all operations except \textit{groupby}. Both \textit{Pandas} and \textit{Dask} have a significantly lower energy consumption for \textit{groupby} operation compared to \textit{Vaex}.
For the dataset \textit{D2}, \textit{Vaex} is the most energy efficient library for \textit{drop} and \textit{merge} operations. \textit{Dask} consumes the least energy for \textit{sort} operation. Similar to D1, \textit{Vaex} is the most energy-expensive library for \textit{groupby} operation with both other libraries having a significantly lower energy consumption.

We can also observe that despite the increase in the number of rows in \textit{D2}, the energy consumption of \textit{Dask} does not change very significantly except in the case of \textit{merge}. The size of the dataset also does not seem to have much influence on \textit{gropuby} operation on any of the libraries. The energy consumption values for the remaining operations increases with the increase in the dataset size for the \textit{Pandas} library. 

The following are the key highlights from the results on row/column operations
\begin{tcolorbox}[colback=red!5!white,colframe=red!75!black]
  \begin{enumerate}
     \item \textit{Vaex} is the most energy-efficient library for \textit{drop}, \textit{concat}, and \textit{merge} operations. 
     \item \textit{Vaex} is the most energy-expensive framework for groupby operation.
     \item The energy consumption of \textit{Dask} does not change very significantly despite change in the size of the dataset.  
     \item For the dataset \textit{D2}, \textit{Pandas} is the most energy-expensive library for all operations other than groupby. 
    \end{enumerate}
\end{tcolorbox}
\begin{table*}[]
\begin{tabular}{|lllllll|llllll|}
\hline
\multicolumn{7}{|c|}{Adult Dataset}                                                                                                                                                   & \multicolumn{6}{c|}{Drugs Review Dataset}                                                                                                             \\ \hline
\multicolumn{1}{|l|}{}          & \multicolumn{2}{l|}{Pandas}                            & \multicolumn{2}{l|}{Vaex}                              & \multicolumn{2}{l|}{Dask}         & \multicolumn{2}{l|}{Pandas}                              & \multicolumn{2}{l|}{Vaex}                              & \multicolumn{2}{l|}{Dask}         \\ \hline
\multicolumn{1}{|l|}{Operation} & \multicolumn{1}{l|}{PKG}   & \multicolumn{1}{l|}{DRAM} & \multicolumn{1}{l|}{PKG}   & \multicolumn{1}{l|}{DRAM} & \multicolumn{1}{l|}{PKG}   & DRAM & \multicolumn{1}{l|}{PKG}    & \multicolumn{1}{l|}{DRAM}  & \multicolumn{1}{l|}{PKG}   & \multicolumn{1}{l|}{DRAM} & \multicolumn{1}{l|}{PKG}   & DRAM \\ \hline
\multicolumn{1}{|l|}{drop}      & \multicolumn{1}{l|}{27.8}  & \multicolumn{1}{l|}{3.6}  & \multicolumn{1}{l|}{12}    & \multicolumn{1}{l|}{0.8}  & \multicolumn{1}{l|}{44.3}  & 3.3  & \multicolumn{1}{l|}{93.5}   & \multicolumn{1}{l|}{20}    & \multicolumn{1}{l|}{6.3}   & \multicolumn{1}{l|}{0.6}  & \multicolumn{1}{l|}{33.6}  & 2.5  \\ \hline
\multicolumn{1}{|l|}{groupby}   & \multicolumn{1}{l|}{3.3}   & \multicolumn{1}{l|}{0.5}  & \multicolumn{1}{l|}{245.7} & \multicolumn{1}{l|}{26.9} & \multicolumn{1}{l|}{5.1}   & 0.4  & \multicolumn{1}{l|}{5.5}    & \multicolumn{1}{l|}{0.5}   & \multicolumn{1}{l|}{272.2} & \multicolumn{1}{l|}{35.3} & \multicolumn{1}{l|}{3.5}   & 0.3  \\ \hline
\multicolumn{1}{|l|}{concat}    & \multicolumn{1}{l|}{102.2} & \multicolumn{1}{l|}{15.5} & \multicolumn{1}{l|}{27.8}  & \multicolumn{1}{l|}{2.8}  & \multicolumn{1}{l|}{139.9} & 9.8  & \multicolumn{1}{l|}{462.1}  & \multicolumn{1}{l|}{92.1}  & \multicolumn{1}{l|}{16.7}  & \multicolumn{1}{l|}{1.6}  & \multicolumn{1}{l|}{105.3} & 7.4  \\ \hline
\multicolumn{1}{|l|}{sort}      & \multicolumn{1}{l|}{57.6}  & \multicolumn{1}{l|}{7.2}  & \multicolumn{1}{l|}{30.1}  & \multicolumn{1}{l|}{2.4}  & \multicolumn{1}{l|}{38.5}  & 2.9  & \multicolumn{1}{l|}{303.8}  & \multicolumn{1}{l|}{48.3}  & \multicolumn{1}{l|}{92.1}  & \multicolumn{1}{l|}{8.7}  & \multicolumn{1}{l|}{30.6}  & 2.2  \\ \hline
\multicolumn{1}{|l|}{merge}     & \multicolumn{1}{l|}{498.5} & \multicolumn{1}{l|}{43.9} & \multicolumn{1}{l|}{46.5}  & \multicolumn{1}{l|}{3.4}  & \multicolumn{1}{l|}{257}   & 17.6 & \multicolumn{1}{l|}{4249.3} & \multicolumn{1}{l|}{534.6} & \multicolumn{1}{l|}{24.2}  & \multicolumn{1}{l|}{1.9}  & \multicolumn{1}{l|}{154.4} & 10.4 \\ \hline
\end{tabular}
\caption{Mean values for energy consumption for missing row/column operations. }
\label{tab:table_operations}
\end{table*}
\subsection{Statistical Aggregation Operations}
\begin{figure}
    \centering
    \includegraphics[scale=0.4]{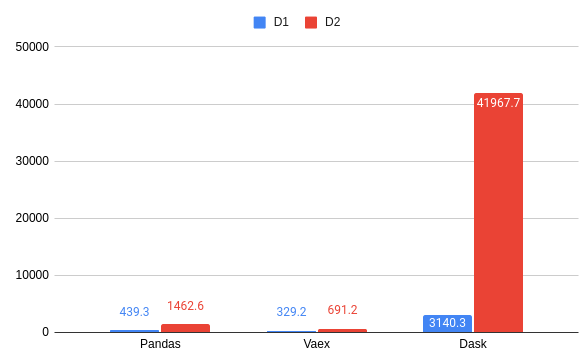}
    \caption{Cumulative energy consumption of statistical aggregation operations of the libraries for both the datasets}
    \label{fig:stat}
\end{figure}
The energy consumption values for the statistical aggregation operations performed on the datasets is shown in Table \ref{tab:statistical_operations}.  For the dataset \textit{D1}, we observe that \textit{Dask} consumes the highest energy for all operations. Of all the operations considered, \textit{count} is the most energy-expensive operation across all libraries. The \textit{sum}, \textit{mean}, \textit{min}, \textit{max} and \textit{unique} are the cheapest in \textit{Pandas} followed by \textit{Vaex}. For the \textit{count}  operation however, \textit{Vaex} consumes the least energy.   

For the dataset \textit{D2}, a similar observation can be made where \textit{Dask} is the most energy-expensive library for all the operations. The \textit{count} operation is the most expensive one for \textit{Pandas} and \textit{Dask}. For \textit{Vaex} however, \textit{unique} is the most energy-expensive operation. Like in case of \textit{D1}, \textit{Pandas} is the least expensive library for all the operations other than \textit{count}. \textit{Vaex} is the least expensive operation for \textit{count} operation. 

The cumulative energy consumption of statistical aggregation operations of the libraries for both the datasets is shown in Figure \ref{fig:stat}. 

The following are the key highlights from the results on statistical aggregation operations. 

\begin{tcolorbox}[colback=red!5!white,colframe=red!75!black]
  \begin{enumerate}
    \item \textit{Dask} is the most expensive library for statistical aggregation operations on both datasets. 
    \item \textit{Pandas} is the most energy-efficient library for all operations considered other than \textit{count}. 
    \item \textit{Vaex} is the most energy-efficient library for \textit{count} operation. 
  \end{enumerate}
\end{tcolorbox}

\begin{table*}[]
\begin{tabular}{|lllllll|llllll|}
\hline
\multicolumn{7}{|c|}{Adult Dataset}                                                                                                                                                   & \multicolumn{6}{c|}{Drugs Review Dataset}                                                                                                                \\ \hline
\multicolumn{1}{|l|}{}          & \multicolumn{2}{l|}{Pandas}                            & \multicolumn{2}{l|}{Vaex}                             & \multicolumn{2}{l|}{Dask}          & \multicolumn{2}{l|}{Pandas}                             & \multicolumn{2}{l|}{Vaex}                              & \multicolumn{2}{l|}{Dask}             \\ \hline
\multicolumn{1}{|l|}{Operation} & \multicolumn{1}{l|}{PKG}   & \multicolumn{1}{l|}{DRAM} & \multicolumn{1}{l|}{PKG}  & \multicolumn{1}{l|}{DRAM} & \multicolumn{1}{l|}{PKG}   & DRAM  & \multicolumn{1}{l|}{PKG}    & \multicolumn{1}{l|}{DRAM} & \multicolumn{1}{l|}{PKG}   & \multicolumn{1}{l|}{DRAM} & \multicolumn{1}{l|}{PKG}     & DRAM   \\ \hline
\multicolumn{1}{|l|}{count}     & \multicolumn{1}{l|}{393.7} & \multicolumn{1}{l|}{25.8} & \multicolumn{1}{l|}{77.2} & \multicolumn{1}{l|}{6.2}  & \multicolumn{1}{l|}{1075}  & 100.5 & \multicolumn{1}{l|}{1218.5} & \multicolumn{1}{l|}{84.4} & \multicolumn{1}{l|}{33.4}  & \multicolumn{1}{l|}{2.6}  & \multicolumn{1}{l|}{11816.3} & 1224.6 \\ \hline
\multicolumn{1}{|l|}{sum}       & \multicolumn{1}{l|}{5.6}   & \multicolumn{1}{l|}{0.4}  & \multicolumn{1}{l|}{34.6} & \multicolumn{1}{l|}{3.3}  & \multicolumn{1}{l|}{379.5} & 40.4  & \multicolumn{1}{l|}{7.5}    & \multicolumn{1}{l|}{0.8}  & \multicolumn{1}{l|}{35.2}  & \multicolumn{1}{l|}{3.2}  & \multicolumn{1}{l|}{5222}    & 551.4  \\ \hline
\multicolumn{1}{|l|}{mean}      & \multicolumn{1}{l|}{2.4}   & \multicolumn{1}{l|}{0.1}  & \multicolumn{1}{l|}{54}   & \multicolumn{1}{l|}{5.3}  & \multicolumn{1}{l|}{380.5} & 40.4  & \multicolumn{1}{l|}{28.7}   & \multicolumn{1}{l|}{3.9}  & \multicolumn{1}{l|}{46.2}  & \multicolumn{1}{l|}{4.4}  & \multicolumn{1}{l|}{5265.4}  & 567.9  \\ \hline
\multicolumn{1}{|l|}{min}       & \multicolumn{1}{l|}{3.3}   & \multicolumn{1}{l|}{0.3}  & \multicolumn{1}{l|}{40}   & \multicolumn{1}{l|}{3.6}  & \multicolumn{1}{l|}{354.9} & 38.9  & \multicolumn{1}{l|}{6.8}    & \multicolumn{1}{l|}{0.6}  & \multicolumn{1}{l|}{50}    & \multicolumn{1}{l|}{5.5}  & \multicolumn{1}{l|}{5185.2}  & 551.8  \\ \hline
\multicolumn{1}{|l|}{max}       & \multicolumn{1}{l|}{3.4}   & \multicolumn{1}{l|}{0.3}  & \multicolumn{1}{l|}{40.2} & \multicolumn{1}{l|}{3.4}  & \multicolumn{1}{l|}{344.8} & 38.2  & \multicolumn{1}{l|}{4.7}    & \multicolumn{1}{l|}{0.6}  & \multicolumn{1}{l|}{34.5}  & \multicolumn{1}{l|}{3.3}  & \multicolumn{1}{l|}{5175.5}  & 532.1  \\ \hline
\multicolumn{1}{|l|}{unique}    & \multicolumn{1}{l|}{4.2}   & \multicolumn{1}{l|}{0.2}  & \multicolumn{1}{l|}{56.2} & \multicolumn{1}{l|}{5.9}  & \multicolumn{1}{l|}{312}   & 35.2  & \multicolumn{1}{l|}{99.6}   & \multicolumn{1}{l|}{7.3}  & \multicolumn{1}{l|}{407.5} & \multicolumn{1}{l|}{39.3} & \multicolumn{1}{l|}{5312.9}  & 562.6  \\ \hline
\end{tabular}
\caption{Mean values for energy consumption for missing statistical aggregation operations. }
\label{tab:statistical_operations}
\end{table*}
\section{Threats to Validity}
\label{sec:threats}
In this section, we discuss potential systematic errors in our work that may pose threats to the outcome of our study. 

One of the most significant threats to the validity of our study are the possible impact of noise, voltage spikes, daemons and other background processes that  may cause inaccurate energy measurement We tried to minimize this issue by repeating the same task 10 times and averaging the recorded values. We also made sure to have the system idle for 30 seconds before the start of the next task to avoid the impact of tail states. Despite the best care taken to ensure minimal impact of these factors, eliminating their effect completely is very difficult as it would require us to have control over the background operations that are needed to run the operating system.Due to hardware constraints, we used a system with basic configuration to run our experiments. The possibility of obtaining significantly different values on a system with a different configuration must also be considered. 

Another factor that may influence the results of our study is the choice of datasets. Libraries such as \textit{Vaex} and \textit{Dask} are optimized for computations and memory usage. The optimizations may influence the amount of work done for performing a certain operation based on the nature of attributes, type and size of the data in the dataset thereby influencing energy readings. To minimize its influence, we performed our experiments with two datasets of different of varying size and attributes. 

The results that we obtained were from the time we set up the study. Open source libraries are often developed by people across the globe. The "competition" between different libraries would mean that the implementations would get more efficient over time. Thus, the results of the experiment obtained in the future for similar experiments may vary based on the modifications to the libraries. Keeping this in mind, we have provided the versions of the libraries used in Section \ref{sec:procedure}. 


\section{Discussion}
\label{sec:discussion}
This paper is our first step towards exploring energy efficiency of the data oriented stages of the machine learning pipeline. Through this work, we also aim to call the attention of the research community to work towards energy efficiency of the data oriented stages of the machine learning pipeline. The results of the study indicate that the choice of libraries for a given task and the format of data storage could potentially influence the energy consumption of the dataframe processing. As an example, based on the results on statistical aggregation operations, we can see that using \textit{Dask} can be a poor choice in terms of energy when aggregation operations are performed very frequently. Similarly, when we have too many and too large dataframes to read, \textit{Dask} can be the most energy-efficient choice. 

Although we have explored the energy consumption of different libraries on a set of tasks, it must be noted that there may exist an alternate ways or functions to accomplish the same task in a library. As an example, many of the missing data handling operations could also be performed using the \textit{apply}\footnote{\url{https://pandas.pydata.org/docs/reference/api/pandas.DataFrame.apply.html}} function. The energy consumption of these alternate methods may be different and they may be more energy-efficient or energy-hungry. However, the current work does not focus on such cases. 

We also observe in some cases that the energy consumption values of a task on a larger dataset (\textit{D2}) is lesser compared to the values for the smaller dataset (\textit{D1}). Few examples of this include the \textit{merge} operation for \textit{Vaex} and \textit{Dask}, \textit{drop} operation for \textit{Dask}, and \textit{concat} operation for \textit{Vaex}. Although it may look counter-intutive, we argue a possible reason for this may be because of other factors such as the number of attributes or the inner working of these libraries. The influence of such factors could be explored further. 

\section{Related Work}
\label{sec:related_work}
In recent years, the energy consumption of software elements have caught the attention of researchers. This has led the researchers to explore and study the energy consumption of software components across various domains with the intention gaining insights that help the developers make energy efficient choices. Measuring the energy consumption and comparing the variations for different available options in the is a popular theme among such studies.

Rouvoy et al. \cite{rouvoy2021comparing} performed an empirical investigation of differences in energy consumption of read/write methods of some of the famous Java libraries. They found some methods to be more efficient than others. Hasan et al. \cite{hasan2016energy} studied the energy profiles of various Java collection classes and found that choice of an inefficient collection can cost as much as 300\% more energy. Schuler et al. \cite{schuler2020characterizing} studied the relation  between System API utilization and energy consumption
in third-party software libraries. Ournani et al. \cite{ournani2021comparing} compared the energy consumption of 27 Java I/O methods for different file sizes and found that the energy consumption varies across APIs with some of them consuming about 30\% less energy than the others. Pereira et al. \cite{pereira2017energy} analysed and compared the energy efficiency across 27 different programming languages. They also explored how energy consumption relates to speed and memory utilization in these languages \cite{pereira2017energy}. Kumar et al. \cite{kumar2019energy} analyzed the energy consumption of Java command line options and found Oracle JDK to be more energy efficient than Open JDK. They also found that UseG1GC and Xint were the most and the least energy efficient command line options respectively \cite{kumar2019energy}. Maleki et al. \cite{maleki2017understanding} studied the impact of OOP design patterns on energy efficiency and found that the use of \textit{overloading} and \textit{decorator} patterns can degrade the energy efficiency of the application. 

Energy efficiency is of a critical importance in battery powered devices such as mobile phones. To this end, exploratory and comparative studies directed towards energy efficiency have also been performed in this area \cite{gupta2011detecting, aggarwal2014power, chowdhury2016client, chowdhury2018exploratory, cruz2019energy, sahin2016does}. 
Gupta et al. \cite{gupta2011detecting} explored the energy consumption of software modules and the impact of co-occurrences of modules on the change in energy consumption values on a windows phone. They also explored the anomalies in power traces. Aggarwal et al. \cite{aggarwal2014power} investigated the relationship between the system call invocation and energy consumption across multiple versions of Android application. Chowdhury et al. \cite{chowdhury2016client} compared the energy consumption of HTTP/2 with its predecessor HTTP/1.1 for mobile apps found that the former outperforms the latter in most scenarios. Chowdhury et al. \cite{chowdhury2018exploratory} studied the energy impact of logging on Android applications and found that limited logging has little to no impact on energy consumption. Cruz et al. \cite{cruz2019energy} analyzed the energy consumption of eight UI automation frameworks and found that certain frameworks can increase the energy consumption by over 2000 percent. Sahin et al. \cite{sahin2016does} explored the energy impact of code obfuscation by studying the impact of 18 obfuscations for 21 usage scenarios accross 11 Android applications on four different mobile phone platforms. They found that obfuscations have a significant impact on energy consumption and is more likely to increase energy consumption than decrease it \cite{sahin2016does}. 

Energy efficiency is crucial in server systems and cloud applications due to its economic and ecological costs. Due to this, many of the-efficiency energy studies have been directed towards this domain \cite{singh2015impact, procaccianti2016empirical, khomh2018understanding}. 
Singh et al. \cite{singh2015impact} explored the energy costs of running several Java APIs on servers to accomplish a set of tasks and found that the developers can reduce the energy costs of server by choosing energy-efficient APIs. Procaccianti et al. \cite{procaccianti2016empirical} performed an empirical investigation of two energy efficient software practices namely "sleep" and "use of efficient queries" and found that they can reduce energy consumption by upto 25\%. Khomh et al. \cite{khomh2018understanding} explored the energy consumption of 6 cloud patterns and found that they can be energy efficient in some cases. 

In the domain of machine learning, several researchers have been working towards improving the energy efficiency. However, the methods mainly remain focused on the machine learning model itself. The methods include model compression techniques like quantization \cite{lee2017lognet, moons2016energy}, pruning \cite{yang2017designing}, distillation \cite{sanh2019distilbert}, use of efficient hardware accelerators \cite{chen2016eyeriss, park2018energy, ko2017design} and use of efficient computation methods in the hardware \cite{sarwar2018energy, jiao2018energy}.  However, the stages of the machine learning pipeline that focus on the data preprocessing and cleaning have largely been ignored. Our exploratory study serves a step in that direction. 

\section{Conclusion and Future Work}
\label{sec:conclusion_future}
In this paper, we presented an exploratory analysis of energy consumption of dataframe processing libraries. This work is intended to be our first study towards the energy efficiency in the data oriented stages of the machine learning pipeline which involves data preprocessing, data cleaning and exploratory data analysis.

We analyzed the energy consumption of three different dataframe processing libraries, namely \textit{Pandas}, \textit{Vaex} and \textit{Dask} for different dataframe processing tasks. The tasks included i/o operations, handling missing data, row/column operations and statistical aggregation operations as shown in Table \ref{tab:tasks}. We used the Intel's RAPL interface to measure the energy consumption of the operations using experimental setttings described in Section \ref{sec:procedure}. We used two datasets from the UCI repository, namely \textit{Adult} Dataset and the \textit{Drug Review} dataset to run our experiments. The results of our experiments contained some interesting observations summarized in Section \ref{sec:results}. The results show us that the choice of library and the format of the data storage can indeed influence the energy consumption of the data preprocessing and data cleaning stages. This choice may be based on factors such as the type of operations performed more frequently, the format in which data is available and the size of the dataset. 

An additional goal of this work is to also call the attention of the research community to focus on the energy efficiency in the stages of the machine learning pipeline other than model training and inferencing. A more comprehensive analysis of energy consumption of the data oriented stages could potentially lead to the discovery of energy efficient practices in data preprocessing, data cleaning and exploratory analysis stages of the pipeline. We have merely scratched the surface of this area and we plan several extensions to the study. In the future, we plan to further explore the energy consumption of dataframe processing libraries on multiple systems with different configurations to get more generalized results. We also plan to repeat the experiments on several larger datasets to explore trends with respect to the energy consumption values for different tasks. The influence of the data types in the dataframe on energy could also be a possible future extension. The list of tasks considered in the study is limited. Given that the libraries have many more functionalities, their energy consumption could also be explored in the future.


\bibliographystyle{ACM-Reference-Format}
\bibliography{references}
\end{document}